\documentclass[aps,prd,twocolumn,preprintnumbers,amsmath,amssymb,superscriptaddress,groupedaddress,floatfix,nofootinbib]{revtex4}

\usepackage[T1]{fontenc} 

\usepackage{graphicx}
\usepackage{ulem}
\usepackage{subcaption}
\usepackage{amsmath} 
\usepackage{verbatim} 
\usepackage{mathrsfs}
\usepackage{color}
\usepackage{amsfonts}
\usepackage{graphics}
\usepackage{epsfig}
\usepackage{hyperref}
\usepackage{caption}
\newcommand{\nk}{\vec{k}}
\newcommand{{\nq}}{\vec{q}}

\newcommand{\x}{\textbf{x}}

\newcommand{\mH}{\mathcal{H}}
\newcommand{\mP}{\mathcal{P}}
\newcommand{\mR}{\mathcal{R}}

\newcommand{\nn}{\nonumber \\}

\newcommand{\dvarphi}{\delta \varphi}

  \newcommand{\avg}[1]{\mathbb{E} [ {#1} ]}

\def \i{\sqrt{i(i+1)}}

\def \k{\kappa}

\def \({\left(}
\def \){\right)}
\def \[{\left[}
\def \]{\right]}

\usepackage{braket}

\newcommand{\n}{{\nonumber}}

\usepackage{braket}
\newcommand{\be}{\begin{equation}}
	\newcommand{\ba}{\begin{eqnarray}}
		\newcommand{\ee}{\end{equation}}
	\newcommand{\ea}{\end{eqnarray}}
\DeclareFontFamily{U}{rsfs}{}         
\DeclareFontShape{U}{rsfs}{m}{n}{<5> rsfs5 <6><7> rsfs7          %
	<8><9><10><10.95><12><14.4><17.28><20.74><24.88> rsfs10}{}     %
\DeclareMathAlphabet{\mathfs}{U}{rsfs}{m}{n}                     %
\newcommand{\mfs}[1]{\mathfs {#1}}

\newcommand{\sO}{{\mfs O}}

\begin{document}

\title{Emergence of cosmic structure from Planckian discreteness}

\author{Gabriel R. Bengochea}
\email{gabriel@iafe.uba.ar}
\affiliation{Instituto de Astronom\'{\i}a y F\'{\i}sica del Espacio (IAFE),
CONICET, Universidad de Buenos Aires, (1428) Buenos Aires, Argentina.}

\author{Gabriel Le\'{o}n}
\email{gleon@fcaglp.unlp.edu.ar}
\affiliation{Grupo de Cosmolog\'{\i}a, Facultad de Ciencias Astron\'omicas y Geof\'{\i}sicas,
Universidad Nacional de La Plata, Paseo del Bosque S/N 1900 La Plata, Argentina.
CONICET, Godoy Cruz 2290, 1425 Ciudad Aut\'onoma de Buenos Aires, Argentina.}

\author{Alejandro Perez}
\email{perez@cpt.univ-mrs.fr}
\affiliation{Aix Marseille Universit\'e, Universit\'e de Toulon, CNRS, Centre de Physique Th\'eorique, 13000 Marseille, France.}

\begin{abstract}

In the standard inflationary paradigm the inhomogeneities observed in the CMB arise from quantum fluctuations of an initially homogeneous and isotropic vacuum state. This picture suffers from two well-known weaknesses. First, it assumes that quantum field theory remains valid at trans-Planckian scales, without modifications from quantum gravity. Second, it necessitates a quantum-to-classical transition in which fluctuations of a homogeneous quantum state become the classical inhomogeneities seen in the CMB. Recently, an alternative paradigm has been proposed in which such inhomogeneities are present from the very beginning, emerging from the assumed discreteness of spacetime at the Planck scale predicted by certain approaches to quantum gravity. Within this framework, scale-invariant scalar perturbations are generated naturally, without relying on trans-Planckian assumptions or invoking a quantum-to-classical transition. Specifically, inhomogeneities in the quantum state at the Planck scale propagate into semiclassical inhomogeneities on CMB scales. Here, we extend the aforementioned proposal to the most realistic case of a quasi-de Sitter expansion; in particular,  we compute the scalar perturbation spectrum as a function of the slow-roll parameters, systematically encoded through the Hubble flow functions.

\end{abstract}

\maketitle
\flushbottom

\section{Introduction}

The inflationary paradigm enjoys substantial support as it addresses several key puzzles in primordial cosmology, including the large-scale homogeneity, isotropy, and spatial flatness of the observable universe. However, these successes are guaranteed only within treatments that effectively truncate the infinitely many physical degrees of freedom; as Penrose has noted \cite{penrose1989difficulties}, serious conceptual difficulties persist. These are related to explaining the extremely special initial conditions required for the second law of thermodynamics to be a feature of our observable universe. From this perspective, the most significant success of the inflationary hypothesis is not its resolution of the so-called flatness and horizon problems, but its prediction of an approximately scale-invariant spectrum of primordial inhomogeneities—remarkably consistent with the cosmic microwave background (CMB) observations \cite{Planck:2018jri}. This prediction relies on the assumption that inhomogeneities in the background fields originate from vacuum fluctuations in the primordial state of geometry and matter.

In this paper, we challenge that view and propose an alternative derivation of the inhomogeneities observed in the CMB, rooted in Planck-scale primordial cosmology. We assume that the ultimate resolution of the flatness, homogeneity, and isotropy problems lies in a yet-unknown aspect of quantum gravity which, like the standard inflationary paradigm, features a de Sitter–like phase of expansion following the Big Bang. We link this hypothesis to observable CMB properties by proposing a tentative mechanism that mitigates another tension in the standard picture: the unexpectedly low tensor-to-scalar ratio revealed by current CMB data. The calculations presented here focus on the generation of scalar perturbations\footnote{Tensor perturbations are deferred to Section \ref{sec_4}, and the quantitative implications will be detailed in a companion work \cite{gabrieles}}.    

Thus, we assume that the observable universe emerges from the Big Bang singularity in a very special initial state, well approximated by a background geometry described by a flat Friedmann-Lemaître-Robertson-Walker (FLRW) metric. In the standard inflationary paradigm, the perturbations of the inflaton, $\dvarphi_k$, are assumed to be in the adiabatic vacuum state which coincides with the Bunch–Davies vacuum in the de Sitter case—$\ket{0}$, for which $\braket{0|\widehat {\delta \varphi_k}|0}=0$. In this state the expectation value of the energy-momentum tensor is homogeneous and isotropic,
\be\label{tandardy}
\braket{0|\widehat T_{ab}(\vec x,t)|0}= \braket{0|\widehat T_{ab}(\vec x+\vec r,t)|0},
\ee
for arbitrary $\vec r$.

However, the state of the universe at later times, when the CMB is produced, is neither homogeneous nor isotropic. This situation requires a symmetry breaking process, often called the quantum-to-classical transition, which raises conceptual tensions related to the measurement problem in quantum cosmology \cite{WIGNER198463, Bell81,Albert,Sudarsky11}. The usual strategy is to adopt one of the familiar positions on that problem (adequate in laboratory settings) and extend it to cosmology. For example, Weinberg \cite{Weinberg:2008zzc} invokes “some form of decoherence”, without specifying the environment in detail; Mukhanov appeals to the many-worlds interpretation \cite{Mukhanov:2005}; collapse theories modify Schrodinger's unitary quantum evolution without observers e.g. \cite{Perez:2005gh,Martin12,Pedro13,Leon16}; others employ a Copenhagen-type measurement scheme, and many do not mention the issue at all. As a consequence, quantum fluctuations in the  adiabatic vacuum are converted into stochastic fluctuations in the observed CMB modes through a process that lies outside the fundamental unitary evolution, which is inherent to quantum mechanics.

This prescription of the traditional inflationary paradigm produces results that agree reasonably well with observations, although the details depend on the specific inflationary model, yet it leaves several conceptual puzzles unresolved \cite{Sudarsky11,Susana13,Bengochea20,Lechuga:2025xlw}. For example, in the case of decoherence, if the universal state is the adiabatic vacuum (or any alternative vacuum; symmetric under the FLRW isometries), no partition into system and environment can produce the symmetry breaking observed in the late time Universe. Under the Copenhagen interpretation, one must assume that an `observation' collapses the initially symmetric quantum state into the inhomogeneous one—manifest in the CMB. This collapse would then lead to structure formation, life on Earth, the founding of the city of Copenhagen, the birth of Bohr and his interpretation, and ultimately the observation of the CMB itself—a retrodiction bringing about the very conditions it seeks to explain.

We recognize that many in the community regard these issues as purely philosophical. This attitude is perhaps understandable after more than fifty years of inflationary modeling, which has accustomed most of us to not asking such questions. After all, we have never needed to answer them when applying quantum theory in the laboratory—why, then, should the Universe be any different? Perhaps another reason is that the apparent empirical success of the standard perspective has fostered the belief that no viable alternative exists. We hope that a humble contribution of this work is, at the very least, to show that alternative accounts are possible. 

Our proposal bypasses the difficult question entirely: the FLRW symmetry is broken from the start as field modes emerge exited away from the vacuum due to their interaction with the microscopic quantum-gravity scale. These excited modes evolve unitarily into the observed CMB anisotropies without invoking decoherence, the measurement problem, or any interpretational subtleties in quantum theory. In this scenario, the primordial state is inhomogeneous from the outset, at the moment when a smooth field-theoretic description first becomes applicable. An early version of this idea was explored in \cite{Amadei:2021aqd} (see also \cite{Leon2022,Pia2023} for related works). Here, we present it in full generality. In particular, we extend its applicability to the slow-roll inflationary context—using the conventional notation and definitions common in the inflationary literature.

Another appealing feature of the model is that it reproduces the CMB observations by invoking a Planck-scale hypothesis, potentially interesting for the insights it may offer into quantum gravity. This same hypothesis naturally yields a low tensor-to-scalar ratio, thereby allowing the initial inflationary epoch to be driven by a much higher (and arguably more natural) value of the primordial vacuum energy. This outcome contrasts with the usual expectation of an inflationary scale far below the Planck scale and may give rise to novel phenomenology that is ruled out in the standard paradigm \cite{Bengochea:2024msf, Christodoulou:2023hyt}.

Before presenting the basic idea in detail, we first examine an assumption in the standard account that has attracted considerable attention in the literature and is commonly referred to as the trans-Planckian problem. This issue concerns the choice of the primordial inflaton state as the adiabatic vacuum or alike \footnote{ We note that a dynamical cosmological background does not admit a uniquely preferred vacuum state. The Bunch–Davies vacuum is simply the natural choice when the background is de Sitter. More generally, different vacuum prescriptions differ mainly in their infrared properties; however, these ambiguities do not affect the standard inflationary mechanism for structure formation, which is driven fundamentally by ultraviolet quantum fluctuations.} (in the idealized case of exact de Sitter inflation). Why, among all possible states in the Hilbert space of the inflaton, should this particular vacuum be selected? The physical idea is that any reasonable deviation from this state would be rapidly diluted by the exponential background expansion. Since the quantum fluctuations relevant for the observed CMB modes originate from extremely short-wavelength modes, the adiabatic vacuum (defined as the asymptotic vacuum in the infinitely distant past) seems a natural idealization of the physical situation. This view can be embraced, only insofar as the problem is analyzed within the framework of quantum field theory in curved spacetime and it is taken for granted that the past singularity does not generate ultra-short-wavelength excitations.

One then finds that the modes responsible for the observed CMB temperature anisotropies originated at wavelengths much shorter than the Planck length during the inflationary epoch. Consequently, it must be assumed that the principles of quantum field theory in curved spacetime remain valid in this trans-Planckian regime. Confidence in this assumption is bolstered by an analogous situation in black-hole evaporation: the thermal Hawking spectrum emerges from the premise that the quantum state of matter approximates the vacuum for modes with trans-Planckian wavelengths near the horizon. Potential deviations due to Planck-scale physics have been studied extensively; several analyses show that even drastic modifications (such as a sharp cutoff or underlying granularity) lead to unobserved alterations in the Hawking spectrum \cite{Jacobson:1999ay, Agullo:2009wt, Baranov:2024myo}. It is therefore expected that a similar robustness characterizes the inflationary case, where similar analyses have been proposed \cite{Martin:2000xs, Niemeyer:2001qe}. However, the previous arguments hold only when the black-hole mass is large or, in the case of de Sitter inflation, when the Hubble rate is small in Planck units. Once these conditions are violated, the description must be replaced by a full quantum-gravity treatment. In such extreme cases, the usual claim that the adiabatic vacuum is ``natural'' loses any rigorous foundation.\footnote{Although this should not be regarded as a firm prediction but rather as an illustrative example, large deviations from Hawking radiation have been reported in symmetry-reduced (toy) models, which can even render quantum black holes stable \cite{Munch:2022teq}. Possible phenomenological implications are discussed in \cite{Bengochea:2024msf, Christodoulou:2023hyt}.}

Consequently,  we will assume that the universe comes from such an extreme inflationary de Sitter-like era, with a Hubble rate of the order of the Planck scale.  Fluctuations in the value of the vacuum energy driving inflation can be described from the perspective of an emergent effective field theory as a theory of a single scalar field \cite{Cheung:2007st, Piazza:2013coa}. Here we adopt a phenomenological perspective on the nature of vacuum energy, without attempting a fundamental explanation of its origin.  We parametrize deviations from a pure cosmological constant in terms of so called Hubble flow functions, which for slow roll inflation are related to the traditional slow-roll  parameters. The value of the vacuum energy matching the natural Planck scale implies that field modes with wavelengths larger than the Planck scale get frozen by the exponential expansion: according to semiclassical field equations that track the expectation values of the field fluctuations, these modes behave as ultra-damped harmonic oscillators.\footnote{In Section \ref{sec_2} we illustrate this behaviour in the idealized case of de Sitter spacetime, and in Section \ref{sec_3} we extend the analysis to a more realistic scenario.}

We will show that the interaction between field modes and a (postulated) Planck-scale discreteness can be modeled so that, when the modes are stretched by the expansion to coincide with the fundamental scale, they are found in a highly excited state. We postulate that, from the effective field theoretic viewpoint, these modes are effectively ``born'' at that instant; thereafter they admit a semiclassical description as their physical wavelengths grow far beyond the Hubble horizon \footnote{ We will describe the mode dynamics by focusing on the evolution of their expectation values which, according to the Ehrenfest theorem, follow the classical field equations. These modes have quantum fluctuations around their mean values which freeze as the modes become super-horizon. A detailed analysis of these fluctuations will be analyzed elsewhere.}. In contrast to the Bunch–Davies  or the adiabatic vacuum, the state of the scalar perturbations that emerges from the fundamental discreteness constitutes a highly excited, inhomogeneous, and anisotropic state. If one were to evolve this state arbitrarily far into the past using standard quantum field theory dynamics, it would appear singular as the excitations would be blue shifted to trans-Planckian scales. However, such an extrapolation is invalid because the effective description necessarily breaks down once the modes approach the fundamental scale again. 

Quantum gravity dynamics would be necessary for these purposes. Once the modes are excited away from the adiabatic vacuum, the admissible states need not respect the background symmetries. Indeed, such symmetry breaking is expected if the familiar smooth field description of quantum field theory emerges from an underlying Planck scale discreteness. In that context, at the fundamental scale there is no well-defined background geometry and the notion of symmetry becomes meaningless (there is no differentiable manifold, no metric, and no Killing fields). Hence, when an effective QFT description first emerges in terms of smooth inflaton modes (as modes stretch into the IR regime) it is natural to expect that these modes would be excited away from the translational invariant state.

Thus, in the present model, the granularity scale is responsible for generating inhomogeneities. This raises a potential concern: the power spectrum of scalar inhomogeneities might retain a memory of the Planck scale, potentially conflicting with the observed approximate scale invariance of the CMB. However, we will show that the inflationary expansion stretches the initial excitations uniformly across all scales, making the predictions of the model consistent with observations for appropriate values of the slow-roll parameters.

The paper is organized as follows. In Section \ref{sec_2} we present the main idea underlying the generation of inhomogeneities and show how the resulting state of the scalar field arises in an idealized, exact de Sitter background. Throughout, we work in the Newtonian gauge and simplify certain equations to highlight key features of the mechanism. Section \ref{sec_3} reviews the linearized gravitational equations that govern the evolution of perturbations on a nearly de Sitter background, characterized by the standard slow-roll parameters. Furthermore, we compute, using the standard gauge invariant (Mukhanov-Sasaki) variable, the power spectrum of inhomogeneities generated by Planck-scale granularity, to leading order in the slow-roll expansion. The predicted features of the associated tensor modes are addressed in Section \ref{sec_4}. Finally, we summarize and discuss our results in Section \ref{discu}. 

Regarding conventions and units, throughout this work, we employ the abstract index notation, which distinguishes between tensor indices and their components. We adopt the spacetime metric signature $(-,+,+,+)$ and work in natural units with $c=\hbar=1$. The Planck mass is denoted by $m_p$, while the \textit{reduced} Planck mass is defined as $M_P \equiv m_p/\sqrt{8 \pi}$.

\section{Generation mechanism in a nut shell} \label{sec_2}

We begin by assuming a spatially flat FLRW metric for the background spacetime, 
\be\label{backflrw}
ds^2=-dt^2+a(t)^2 d\vec x\cdot d\vec x,
\ee
with $a(t)$ the scale factor and $u^a=\partial_t^a$ the four-velocity of comoving observers. Moreover, we will assume that the primordial value of the cosmological constant (or vacuum energy density) has its natural value $\Lambda \approx m_p^2$.  The standard  Hubble parameter is expressed as $H =  \dot a/a$,  the dot denotes the derivative with respect to cosmic time $t$. Consequently, the scale factor evolves as $a(t)=a_0 \exp(H t)$, with the Hubble rate $H \approx m_p$. Such primordial dark energy is expected to decay and produce the reheating of the universe that leads to a hot (post-inflationary) big bang. We are not concerned here with the details of such a reheating phase. In fact, from an effective field theory perspective \cite{Cheung:2007st,Piazza:2013coa}, an evolving dark energy component can be encoded in the dynamics of a single inflaton field $\varphi$ with potential $V(\varphi)$.

The inflationary phase can be characterized by the so-called Hubble flow functions $\epsilon_i$ (HFF) \cite{terreroHFF,terreroHFF2}. These are defined as
\begin{equation}\label{defepsilonn}
	\epsilon_{n+1} \equiv \frac{d \ln \epsilon_n}{d N}, \qquad \epsilon_0 \equiv \frac{H_{\text{ini}}}{H},
\end{equation}
where $N \equiv \ln (a/a_{\text{ini}}) $ is the number of e-folds from the beginning of inflation.
From the previous equations, it follows the expansion 
\be
\log(H_{\rm ini}/H)=\epsilon_1 N +\frac12 \epsilon_1\epsilon_2 N^2 +\frac16(\epsilon_1\epsilon_2^2+\epsilon_1\epsilon_2\epsilon_3) N^3+\sO(N^4).
\ee
Inflation occurs if $\epsilon_1 < 1$ and the slow-roll approximation assumes that all these parameters are small during inflation $|\epsilon_n| \ll 1$. Additionally, since $d N = H dt $, it is straightforward to obtain an equivalent expression for the HFF, i.e.
\begin{equation}\label{dotepsilon}
	 \dot \epsilon_n = H \epsilon_n \epsilon_{n+1}.
\end{equation} 
As exposed in \cite{terreroHFF}, one of the main advantages of definition \eqref{defepsilonn} over other definitions of the traditional slow-roll parameters is that it is a  model independent way to characterize inflation (i.e. not restricted to models with a single scalar field).  Nevertheless, for the particular case of slow-roll inflation, the HFF can be related to the usual slow-roll parameters \cite{Leach:2002ar}.

Returning to our model, during the inflationary regime we assume the potential $V(\varphi)$ to be approximately constant, with $V(\varphi)/m_p^4 \approx 1$. Departures from exact constancy are quantified by the HFF; for a pure de Sitter expansion they vanish, $\epsilon_n = 0$ for all $n$. We interpret this scalar-field theory as an effective description of the collective mode emerging from the microscopic degrees of freedom that generate the cosmological constant, with $V(\varphi)$ setting the evolving value $\Lambda = V(\varphi)/M_P^2$.
\footnote{Alternatively, in the formulation of Refs.~\cite{Perez:2017krv,Perez:2018wlo,Amadei:2021aqd}, one has $8\pi \nabla^a T_{ab}(\varphi)=m_p^2 \nabla_b \Lambda$, where $T_{ab}(\varphi)$ is the energy–momentum tensor of the free scalar field.}

Next, we separate the scalar field into an homogeneous part plus some small perturbations $\varphi=\varphi_0+\delta\varphi$, and we assume that the quantum state describing the scalar field perturbations 
$\ket{ \Psi }  \neq \ket{0}$, where $\ket{0}$ is any vacuum state (e.g. the Bunch-Davies vacuum in the case of de Sitter or the adiabatic vacuum in general) for which the expectation value of the field is symmetric\footnote{{In what follows, do not confuse the state $\ket{\Psi}$ with the scalar perturbation to the metric $\Psi$ in the Newtonian gauge.}}. In fact, we define the (non vanishing) mean field variable 
\be\label{semi}
\delta \varphi_k\equiv \braket{\Psi|\widehat {\delta \varphi_k}|\Psi}\not=0, \ \ \ k\le m_p a
\ee
where we are expressing the previous equation in terms of Fourier modes which are related to the actual field perturbations via 
\be
\delta\varphi(t, \vec x)=\frac{1}{(2\pi)^{\frac32}} \int d^3k \: \delta \varphi_{\vec k}(t) \exp(i \vec k\cdot \vec x),
\ee
with reality conditions
\be
\overline{\delta \varphi_{\vec k}}=\delta \varphi_{-\vec k}.
\ee

The mechanism responsible for producing these excitations away from the vacuum, arising from Planck scale discreteness, was outlined in the previous section. A key distinction from the standard paradigm is that the expectation value of the energy–momentum tensor evaluated in the quantum state of the scalar field already breaks homogeneity at the primordial level. In fact due to \eqref{semi}, and in contrast with the standard choice \eqref{tandardy}, we have that 
\be
\braket{\Psi|T_{ab}(\vec x, t)|\Psi}\not=\braket{\Psi|T_{ab}(\vec x+\vec r, t)|\Psi}.
\ee
Thus, the state $\ket{ \Psi }$ characterizes an inhomogeneous and anisotropic energy-momentum tensor, and because of Einstein semiclassical equations $G_{ab} = 8 \pi G \langle T_{ab} \rangle$, the spacetime also departs from perfect spatial symmetry, i.e. the ``primordial curvature perturbations'' are sourced by $\ket{ \Psi }$.

We now focus on the dynamics of the perturbations and its generation mechanism. First, we will assume that the expectation value of the scalar field \eqref{semi} satisfies the classical field equations for modes $k\ll m_p a$, (in the Newtonian gauge) namely 
\begin{eqnarray}\label{km}
 & & \delta \ddot\varphi_{\vec k}+3 H \delta\dot \varphi_{\vec k}+\frac{k^2}{a^2} \delta \varphi_{\vec k}+\frac{d^2V(\varphi_0)}{d \varphi^2} \delta \varphi_{\vec k} \nn 
 &+&2\frac{dV(\varphi_0)}{d \varphi} \Psi_k -4\dot\varphi_0\dot \Psi_k=0,
\end{eqnarray}
after created by the interaction with the granularity at the Planck scale and being stretched to long super-horizon scales. 

On the other hand, the very notion of mode and its dynamics when $k \approx a m_p$ belong to the domain of quantum gravity and is therefore poorly understood. Here, we adopt the viewpoint that the underlying spacetime granularity excites the modes such that, once they become super-Hubble ($k \ll a m_p$) they admit a semiclassical description. However, the generation mechanism is quantum-gravitational in nature, {therefore} we model our lack of precise knowledge by a Brownian stochastic process. Such model should capture the quantum gravity process in stochastic averages faithfully if many degrees of freedom are involved in a sufficiently noisy fashion.
Specifically, we add a noise source term to the field equation \eqref{km} and extend its applicability up to $k = a m_p$; in this manner, the stochastic source provides a statistical representation of the quantum gravitational effects, namely,
\begin{eqnarray}\label{newkm}
 & & \delta \ddot\varphi_{\vec k}+3 H \delta\dot \varphi_{\vec k}+\frac{k^2}{a^2} \delta \varphi_{\vec k}+\frac{d^2V(\varphi_0)}{d \varphi^2} \delta \varphi_{\vec k} \nn 
 &+&2\frac{dV(\varphi_0)}{d \varphi} \Psi_k -4\dot\varphi_0\dot \Psi_k
 =\xi_{\vec k}(t-t^{\rm HC}_k), 
\end{eqnarray}
where $\xi_{\vec k}(t - t^{\rm HC}_k)$ is a stochastic source peaked at zero, and $t^{\rm HC}_k$ denotes the horizon-crossing (HC) time. This construction idealizes the notion that the granular structure, acting at time $t = t^{\rm HC}_k$ and at the fixed physical scale $H \approx m_p$, interacts with the scalar field precisely when the mode's wavelength becomes Planckian, i.e. when $k = a(t^{\rm HC}_k) m_p$.

In order to attain an intuitive understanding of our proposed mechanism, let us consider the simple case in which
\be\label{source}
\xi_{\vec k}(t-t^{\rm HC}_k)=\Delta_{\vec k} \delta(t-t^{\rm HC}_k),
\ee
where the stochasticity has been absorbed into the variable $\Delta_{\vec k}$. The random nature of $\Delta_{\vec k}$ is encoded in the ensemble average, denoted by $\avg{.}$, through the conditions
\be\label{primero}
\avg{\Delta_{\vec k}}=0,
\ee
and
\be\label{segundo}
\avg{\Delta_{\vec k}\Delta_{\vec q}}=P_{\Delta}(k) \delta(\vec k+\vec q),
\ee
where the structure of the second moment is fixed by translational invariance, and $P_{\Delta}(k)$ is the power spectrum of the sources.

Before performing any statistical analysis, we integrate Eq. \eqref{newkm} across the interaction time $t^{\rm HC}_k$ over an infinitesimally small interval. We obtain
 \be
 \Delta \delta\dot\varphi_{\vec k}+3H\Delta \delta\varphi_{\vec k}=\Delta_{\vec k} ,
 \ee
where $\Delta \delta\dot\varphi_{\vec k}$ and $\Delta \delta \varphi_{\vec k}$ encode the change in the field degrees of freedom around horizon crossing due to the source \eqref{source}. We must determine how the interaction ``kicks'' the field; in other words, how the perturbation is distributed between $\delta\dot\varphi_{\vec k}$ and $\delta\varphi_{\vec k}$. By analogy with Brownian diffusion, and for concreteness, we can assume the ``equipartition''
 \be\label{BE}
 \Delta \delta\dot\varphi_{\vec k}=H \Delta \delta\varphi_{\vec k},
  \ee
which leads to the following initial condition for the modes at horizon crossing:
 \be
 \left. \Delta \delta\varphi_{\vec k}\right|_{k=am_p}=\frac{\Delta_{\vec k}}{4H}.
 \ee
This equipartition assumption is introduced solely for illustrative purposes; as we show below, it is not required for predicting the impact of the stochastic sources on the scalar-perturbation power spectrum observed in the CMB.
 
For super-Hubble modes, $k^2 < a^2 H^2$, the friction term dominates and the solutions freeze exponentially once they cross the horizon. The preceding model thus provides a simple realization of the stochastic production of inhomogeneities, in which scalar-field modes are generated by the random source $\Delta_{\vec k}$, acting in a delta function like manner at horizon crossing. However, until this point we have not yet defined the stochastic process completely. The main idea will be illustrated in a simplified context in the following sub Section. The definition in full generality is given in Section \ref{sec_3}.

\subsection{Fluctuation dissipation and Einstein balance in the idealized de Sitter case}\label{subsecDS}

We now focus on presenting our proposed mechanism, in the idealized case of a perfect de Sitter spacetime. Hence, we assume that $H$ and $d^2V(\varphi_0)/d\varphi_0^2$ remain strictly constant. Discarding the $k^2/a^2 \delta\varphi_{\vec k}$ term in the field equation \eqref{newkm}, which rapidly vanishes as the modes exit the horizon, and neglecting the scalar gravitational contributions (absent in exact de Sitter spacetime \cite{PeterUzan:2009}), we obtain the simplified dynamical equation
\be\label{eoml}
 \delta \ddot\varphi_{\vec k}+3H \delta\dot\varphi_{\vec k}=\xi_{\vec k}(t-t^{\rm HC}_k). 
\ee

Equation \eqref{eoml} is the exact analog of a free-particle Langevin equation for each mode, driven by a stochastic source $\xi_{\vec k}(t-t^{\rm HC}_k)$, whose late-time asymptotic solution is given by
\ba\label{solution}
\delta\varphi_{\vec k}(t)&=&  \int_0^t dt'  G(t-t') \xi_{\vec k}(t'-t^{\rm HC}_k) ,
\ea 
where the Green's function is
\ba\label{green}
G(t-t')&\equiv& \exp[{-3H (t-t')}] \frac{\sinh\left[3H (t-t')\right]}{3H} .\ea 
From the Eq. \eqref{solution} we also obtain 
\ba\label{solutiondot}
&& \delta\dot\varphi_{\vec k}(t)=  \int_0^t dt'  \dot G(t-t') \xi_{\vec k}(t'-t^{\rm HC}_k). 
\ea 

The energetic cost associated with producing inhomogeneities is central to the generation mechanism we propose. Assuming that no source-free modes are present initially, Eq. \eqref{primero} implies $\avg{ \delta\varphi}  = 0$. Consequently, a meaningful definition of the average energy must include the quadratic contributions of the perturbations to the energy–momentum tensor. Hence, we define 
\be
\rho^{(2)}(t,\vec x)=\frac{1}{2}\delta\dot \varphi^2(t,\vec x)
\ee
and 
\be
P^{(2)}(t,\vec x)=\frac{1}{2}\delta\dot \varphi^2(t,\vec x)
\ee
The equations of motion \eqref{eoml} imply the continuity equation 
\ba
&&  \dot\rho^{(2)}+3H (\rho^{(2)}+P^{(2)})=\\
&& \n =\frac{1}{8\pi^3}\int d^3 \vec k d^3 \vec q \   \xi_{\vec k}(t-t^{\rm HC}_k) \delta\dot\varphi_{\vec q}(t) \ \exp(i(\vec k+\vec q)\cdot \vec x). \\
&& =\frac{1}{8\pi^3}\int_0^{t} dt_1   \int d^3 \vec k d^3 \vec q \  \dot G(t-t_1) \nn 
&& \times  \Delta_{\vec k} \Delta_{\vec q} \ \delta(t-t^{\rm HC}_k) \delta(t_1-t^{\rm HC}_q) \ \exp(i(\vec k+\vec q)\cdot \vec x),\n 
\ea
which implies that in the absence of sources, $\xi_k=0$, there is energy conservation. The presence of the source $\xi_{\vec k}(t-t^{\rm HC}_k)$ produces a non trivial diffusion work. 
We are interested in the ensemble average of the previous equation. Therefore, using \eqref{segundo}, we obtain
\ba
& & \avg{\dot\rho^{(2)}+3H (\rho^{(2)}+P^{(2)})}= \nn 
& & \frac{1}{2\pi^2} \int k^2 dk \ \dot G(0) P_{\Delta_k}\  \delta(a^{-1}[k/m_p]-t_k^{HC}). 
\ea
where in the second line $a^{-1}[\ \ ]$ denotes the inverse of the function $a(t)$. Finally, using that $\dot G(0)=1$ and evaluating the integral in $dk$ we arrive at the desired result:
\be\label{desired}
\avg{\dot\rho^{(2)}+3H (\rho^{(2)}+P^{(2)})}=\left.\frac{k^3 H P_{\Delta_k}}{2\pi^2} \right|_{k=a H.}
\ee

Let us note that explicit knowledge of the Green function is not required in the previous derivation. The key lesson of the present analysis (which we will show to hold, in general, for small deviations from the present de Sitter idealization) is that the energy injected into the inhomogeneities is a functional of the power spectrum of the stochastic sources. This observation allows us to express the generation mechanism in a model-independent fashion, without relying on specific details of the background dynamics, as we explain in the next section. Notice that the last equation implies that, in order for the energy injected into inhomogeneities to remain constant in time (which is what one would expect during the de Sitter phase), the power spectrum must be $P_{\Delta_k}= P_0/k^3$, i.e., it must be scale invariant. The scale invariance of the power spectrum of scalar perturbations is therefore not tied to any particular property of vacuum fluctuations; rather, it arises because the interaction with the microscopic granular structure of de Sitter spacetime feeds the production of inhomogeneities through a steady energy input.

\subsection{The steady energy flow condition}\label{subsecEF}

In this final part, we generalize the key equations of the previous subsection. The resulting expressions will be applied in Section \ref{sec_3}. Let us consider a scalar field obeying
\be\label{fie0}
\nabla^a\nabla_a \varphi-\partial_\varphi V[\varphi]=\xi,
\ee
where $\xi$ is the stochastic source introduced earlier.
The canonical energy–momentum tensor of the scalar field is  given by
\be
T_{ab}=\nabla_a\varphi\nabla_b\varphi-\frac12 g_{ab} (\nabla_c\varphi\nabla^c\varphi+2V[\varphi]).
\ee
Consequently, the divergence of the energy-momentum tensor is 
\be
\nabla^aT_{ab}=\xi \nabla_b\varphi,
\ee
provided the field equation \eqref{fie0} is satisfied. 

When applied to cosmology, the injection of energy that generates inhomogeneities—via the interaction between Planck-scale granularity and the field modes—can depend only on the Hubble rate, the sole relevant scale. If the Hubble rate is approximately constant, the energy flux is steady. Dimensional analysis then yields
\be\label{ghhhhh}
\avg{\nabla^aT_{ab}}u^b=\avg{\xi \nabla_b\varphi}u^b=\gamma H^5,
\ee
where $u^a$ is the comoving four-velocity (to zeroth order, as higher-order contributions are at least second order) and $\gamma$ is a dimensionless constant. In the FLRW background \eqref{backflrw}, the left-hand side of Eq. \eqref{ghhhhh} coincides with Eq. \eqref{desired}; equating it to the right-hand side, $\gamma H^5$, implies that the source power spectrum $P_{\Delta}(k)$ must be scale-invariant, as discussed previously.

\section{The stochastic generation of inhomogeneities from granularity}\label{sec_3}

In this section we will analyze our proposal in the most realistic case of a quasi-de Sitter spacetime. In particular, we will compute the evolution of $\dvarphi$ in a quasi-de Sitter background, when emerging from the Planck scale excited by the interaction with the Planckian microscopic structure. We propose a precise treatment of the simple discussion of Sect. \ref{sec_2}. We will focus only in the scalar sector of the metric-matter perturbations. 

It is convenient to switch to conformal time $\eta$ (i.e. $d \eta = dt/a$) coordinates; hence, the metric components up to first order in the perturbations are given generically by
\begin{subequations}\label{metricaperts}
	\begin{equation}\label{eqg00}
		g_{00} = - a^2(1+2 \phi),
	\end{equation}
	\begin{equation}\label{eq0i}
		g_{0i} = a^2  \partial_i B,
	\end{equation}
	\begin{equation}\label{eqij}
		g_{ij} = a^2 (\delta_{ij} - 2 \psi \delta_{ij} + \partial_{ij} 2 E  ),
	\end{equation}
\end{subequations}
where $\phi$, $\psi$, $B$  and $E$ are scalar functions dependent on the (conformal) time and spatial coordinates ($i=1,2,3$).

On the other hand, we have modeled matter as the scalar field $\varphi = \varphi_0 + \dvarphi$. As a consequence, the components of the corresponding energy-momentum tensor at first order are 
\begin{subequations}
	\begin{equation}
		\delta T^0_0 = -\frac{1}{a^2} (\varphi_0' \dvarphi'-\varphi_0'^2 \phi)- \partial_\varphi V \dvarphi
	\end{equation}
	\begin{equation}
		\delta T^0_i = -\frac{1}{a^2} (\varphi_0' \dvarphi,_i)
	\end{equation}
	\begin{equation}
		\delta T^i_j = \frac{1}{a^2}   (\varphi_0' \dvarphi'-\varphi_0'^2 \phi-a^2 \partial_\varphi V \dvarphi) \delta^i_j
	\end{equation}
\end{subequations}
where a prime denotes derivative with respect to $\eta$. We can now write the Einstein field equations (EFE) at first order in the perturbations,
\begin{subequations}\label{dEFE}
	\begin{eqnarray}\label{EFE00}
	&&	3\mH (\mH \phi + \psi') - \nabla^2 (\psi + \mH \sigma) \nn 
	&&=  4 \pi G ( -\varphi_0' \dvarphi'+ \varphi_0'^2 \phi - a^2 \partial_\varphi V \dvarphi ) ,
	\end{eqnarray}
	\begin{equation}\label{EFE0i}
		\mH \phi + \psi' = 4 \pi G \varphi_0' \dvarphi,
	\end{equation}
	\begin{equation}\label{EFEij}
		\sigma' + 2\mH \sigma + \psi - \phi = 0,
	\end{equation}
	\begin{eqnarray}\label{EFEii}
		&& \psi'' + 2\mH \psi' + \mH \phi' + ( 2\mH' + \mH^2) \phi \nn 
		&&= 4 \pi G (\varphi_0' \dvarphi'-\varphi_0'^2 \phi-a^2 \partial_\varphi V \dvarphi).
	\end{eqnarray}
\end{subequations}
We have introduced the definitions $\mH \equiv a'/a$, $\sigma \equiv E'-B$.  Note that Eqs. \eqref{dEFE}, are expressed in an arbitrary gauge. The first-order scalar field perturbations obey the wave equation
\begin{eqnarray}\label{dphievogen}
	&& \dvarphi'' + 2\mH \dvarphi' - \nabla^2 \dvarphi + a^2 \partial^2_{\varphi \varphi} V \dvarphi \nn 
	&& = -2 a^2 \partial_\varphi V \phi + \varphi_0' [\phi' + 3 \psi' - \nabla^2 \sigma ],
\end{eqnarray}
where the terms on the right-hand-side represent the effect of metric perturbations at first-order (sometimes called the gravitational back-reaction).

Let us note that, in the Newtonian gauge, $B=E=0$. Then, $\sigma =0$ and also $\phi=\psi$ from Eq. \eqref{EFEij}. Therefore, in such a gauge, Eq. \eqref{dphievogen} (after switching to Fourier's space) is exactly Eq. \eqref{km} when the latter is expressed in conformal coordinates. 
Furthermore, in Fourier space, we can model the creation of the mode $\dvarphi_k$ via the addition of the source term, in the same manner as we did in the de Sitter case. This is, 
\begin{eqnarray}\label{dphievo2}
	&& \dvarphi_k'' + 2\mH \dvarphi_k' +k^2 \dvarphi_k + a^2 \partial^2_{\varphi \varphi} V \dvarphi_k \nn
	&& + 2 a^2 \partial_\varphi V \phi_k - \varphi_0' [\phi_k' + 3 \psi_k' + k^2 \sigma_k ] \nn 
	&& =  a \Delta_{\nk} \delta(\eta-\eta_k^{\rm HC}),
\end{eqnarray}
where $\eta_k^{\rm HC}$ is the conformal time at horizon crossing for the mode $k$. Once again, in the Newtonian gauge, and changing to cosmic time coordinates $t$, Eq. \eqref{dphievo2} results in Eq. \eqref{newkm}.

The gauge-invariant variable $v$, defined as
\begin{equation}\label{defMS}
	v \equiv a \left( \dvarphi + \frac{\varphi_0'}{\mH} \psi \right),
\end{equation}
also called the Mukhanov-Sasaki (MS) variable, will be a useful quantity for computing the primordial power spectrum. The MS variable is related to another important quantity $\mR$ \cite{Brandenberger:2003vk}, defined as
\begin{equation}
	\mR \equiv \frac{v}{z} \qquad \textrm{with} \qquad z\equiv \frac{\varphi'_0}{\mH}a= \sqrt{2 \epsilon_1} M_P a,
\end{equation}
where $\mR$ is the curvature perturbation in the comoving gauge, and in the last equality we used the exact relation (obtained from the Friedmann equation)
\begin{equation}\label{eqeqeq}
	 \varphi_0'^2 = 2 \epsilon_1 \mathcal{H}^2 M_P^2,
\end{equation}
with $M_P$ the reduced Planck mass. In fact, the primordial power spectrum, which is the standard theoretical prediction of inflation, is expressed as the power spectrum of $\mR$.

In order to find a solution of Eq. \eqref{dphievo2}, we will choose to work in the \textit{spatially flat} gauge, characterized by $\psi = E = 0$. Moreover, in this gauge, the Mukhanov-Sasaki variable (which is a gauge-invariant quantity), is simply $v =  a \dvarphi$ [see \eqref{defMS}]. Therefore, after finding the solution $\dvarphi$ in such a gauge, the primordial power spectrum $\mP_\mR$ can be obtained in a straightforward manner. 

Algebraic manipulations employing EFEs \eqref{dEFE} allow us rewrite Eq. \eqref{dphievo2} in the spatially flat gauge. This results in
\begin{eqnarray}\label{dphievoflat}
	&& \dvarphi_k'' + 2\mH \dvarphi_\k' + k^2 \dvarphi_k  \nn
	& & + \left[ a^2 \partial^2_{\varphi \varphi} V - \frac{8 \pi G}{a^2}  \left(  \frac{a^2 \varphi_0^{'2}}{\mH} \right)'   \right] \dvarphi_k = a \Delta_{\nk} \delta(\eta-\eta_k^{\rm HC}). \nn
\end{eqnarray}
The effect of gravitational coupling is explicitly manifested in the term proportional to $G$. Also, in the exact de Sitter case, this same term vanishes exactly; meanwhile, in the quasi-de Sitter case, that term is proportional to the HFFs as can be seen from Eq. \eqref{eqeqeq}.

Next, we rewrite Eq. \eqref{dphievoflat} in terms of the  Mukhanov-Sasaki variable $v_k = a \dvarphi_k$, which yields
\begin{eqnarray}
	&& v_k'' + \left[k^2 - \frac{a''}{a} + a^2 \partial^2_{\varphi \varphi} V - \frac{8 \pi G}{a^2}  \left(  \frac{a^2 \varphi_0^{'2}}{\mH} \right)'     \right] v_k \nn 
	& & = a^2 \Delta_{\nk} \delta(\eta-\eta_k^{\rm HC}). 
\end{eqnarray}
Expressing the previous equation in terms of $z$, we finally have
\begin{equation}\label{MS2}
		v_k'' + \left[k^2 - \frac{z''}{z}    \right] v_k = a^2 \Delta_{\nk} \delta(\eta-\eta_k^{\rm HC}),
\end{equation}
which is the Mukhanov-Sasaki equation with a the stochastic source. The solution of the above equation is given by
\begin{equation}\label{solvk}
	v_k (\eta) = \int_{-T}^\eta d\tau \: G_k(\eta,\tau) a^2(\tau) \Delta_{\nk} \delta(\tau-\eta_k^{\rm HC}),
\end{equation}
where $-T$ is the conformal time at the beginning of inflation, and $G_k(\eta,\tau)$ is the Green's function of Eq. \eqref{MS2}. In particular, it is given by
\begin{equation}\label{greenvk}
	G_k(\eta,\tau) = \frac{ \Theta(\tau-\eta) v^+_k(\eta) v^-_k (\tau) - \Theta(\eta-\tau) v^-_k(\eta) v^+_k (\tau)}{W(\tau)}.
\end{equation}
The functions $v^{\pm}_k(\eta)$ are two linearly independent solutions of the homogeneous Mukhanov-Sasaki equation
\begin{equation}\label{MS0}
	v^{\pm ''}_k + \left[k^2 - \frac{z''}{z}    \right] v^{\pm}_k = 0,
\end{equation}
and $W(\tau)$ is the Wronskian associated to those solutions, and $\Theta(\tau)$ is the Heaviside function. 
At first order in the HFFs, the solutions $v^{\pm}_k(\eta)$ are
\begin{subequations}
	\begin{equation}
		v^+_k (\eta) = \sqrt{\frac{-\eta \pi}{2}} J_\nu (-k \eta),
	\end{equation}
		\begin{equation}
		v^-_k (\eta) = \sqrt{\frac{-\eta \pi}{2}} Y_\nu (-k \eta),
	\end{equation}
\end{subequations}
with
\begin{equation}\label{nuni}
	\nu \simeq \frac{3}{2} + \epsilon_1 + \frac{\epsilon_2}{2}
\end{equation}
where $J_\nu$, $Y_\nu$ are Bessel functions of the first and second kind respectively of order $\nu$. Moreover, for these solutions we have $W(\tau) = 1$. 

Our primary objective is to obtain the primordial power spectrum. To this end, we employ the insights gained from the simpler case of perfect de Sitter expansion, analyzed in sections \ref{subsecDS} and \ref{subsecEF}.

\subsection{The power spectrum of $\Delta_{\nk}$}

The first task is to obtain the power spectrum of the stochastic variable $\Delta_{\nk}$. Afterward, this result is used to compute the primordial power spectrum, namely $\mP_\mR (k)$. The former task is addressed in the present subsection, and the latter in the following subsection.

Our starting point is equation \eqref{ghhhhh}, in particular, we focus on the second equality
\begin{equation}
	\avg{\xi \nabla_a \varphi} u^a = \gamma H^5.
\end{equation}
In conformal time coordinates, $u^a = a^{-1} (\partial_\eta)^a$; consequently, the previous equation is
\begin{equation} \label{main1}
		\avg{a^{-1}  \xi \dvarphi '(x)} =  \gamma H^5,
\end{equation}
where we have used the expansion $\varphi = \varphi_0 + \dvarphi$ , and that $\avg{a^{-1}  \xi \varphi_0'} = 0$. Expressing the left-hand-side of Eq. \eqref{main1} in Fourier modes, we have
\begin{eqnarray} \label{main2}
	\avg{a^{-1} \xi \dvarphi '(x)} &=&  \int\frac{d^3 k}{(2 \pi)^{3/2} } \int\frac{d^3 q}{(2 \pi)^{3/2} } \avg{ a^{-1}  \dvarphi_k'(\eta) \Delta_{\nq} } \nn
	&\times&  \frac{\delta(\eta-\eta_q^{\rm HC})}{a(\eta)} e^{ i \x \cdot ({\nq} + {\nk})}.
\end{eqnarray}
In the last equation, we have written $\xi$ in conformal time coordinates; i.e., given that $\xi_{\nk} = \Delta_{\nk} \delta (t-t_k^{\rm HC})$, in conformal coordinates this results in $\xi_{\nk} = \Delta_{\nk} \delta (\eta-\eta_k^{\rm HC})/a$.

As mentioned in the previous subsection, in the spatially flat gauge the perturbation $\dvarphi$ is related directly to the MS variable $v$, namely $\dvarphi = v/a$. Working in this gauge, we obtain in Fourier space
\begin{equation}
	\dvarphi'_k = \frac{v_k'}{a} - \frac{\mH}{a} v_k.
\end{equation}
We can now use the solution for $v_k$ (found in Eq. \eqref{solvk}) in the latter equation, which yields 
\begin{eqnarray}\label{dphiprimak}
		\dvarphi_k'(\eta) &=&  \frac{1}{a(\eta)}  \int_{-T}^\eta d\tau \:  a (\tau)^2  \Delta_{\nk} \delta(\tau - \eta_k^{\rm HC}) \nn 
		&\times& \left[ G_k'(\eta,\tau) - \mH(\eta) G_k(\eta,\tau) \right] 
\end{eqnarray}
where $G_k(\eta,\tilde \eta)$ is the Green's function found in Eq. \eqref{greenvk}. 

Substituting Eq. \eqref{dphiprimak} in Eq. \eqref{main2}, we obtain 
\begin{eqnarray} \label{main3}
	\avg{ a^{-1} \xi \dvarphi '(x)} &=& \int\frac{d^3 k}{(2 \pi)^{3/2} } \int\frac{d^3 q}{(2 \pi)^{3/2} }   e^{ i \x \cdot ({\nq} + {\nk})} \frac{\delta(\eta-\eta_q^{\rm HC})}{a(\eta)^3} \nn 
	&\times&  \avg{\Delta_{\nk} \Delta_{\nq}}   \int_{-T}^\eta d\tau \: a (\tau)^2   \delta(\tau - \eta_k^{\rm HC}) \nn
	&\times& \left[ G_k'(\eta,\tau) - \mH(\eta) G_k(\eta,\tau) \right].
\end{eqnarray}
The isotropy and homogeneity of the stochastic process requires
\begin{equation}
\avg{\Delta_{{\nk}} \Delta_{{\nq}}}\equiv P_{\Delta_k} \  \delta^{3}({\nk}+{\nq}).
\end{equation}
Consequently, substituting the previous equation in Eq. \eqref{main3}, and integrating over ${\nq}$ we obtain 
\begin{eqnarray} \label{main4}
	\avg{a^{-1} \xi \dvarphi '(x)} &=& 4 \pi \int\frac{d^3 k}{(2 \pi)^{3} }  \frac{\delta(\eta-\eta_k^{\rm HC})}{a(\eta)^3} P_{\Delta_k}    \nn
	&\times& \int_{-T}^\eta d\tau \: a (\tau)^2   \delta(\tau - \eta_k^{\rm HC}) \nn 
	&\times& \left[ G_k'(\eta,\tau) - \mH(\eta) G_k(\eta,\tau) \right]
\end{eqnarray}
Next, integrating over $\tau$ and the angular part of ${\nk}$ in the previous expression yields,
\begin{eqnarray}\label{main5}
		\avg{a^{-1} \xi \dvarphi '(x)} &=& \frac{2}{\pi } \int dk \:  k^2  \frac{\delta(\eta-\eta_k^{\rm HC})}{a(\eta)^3} P_{\Delta_k} a(\eta_k^{\rm HC})^2 \nn
		&\times& \left[ G_k'(\eta,\eta_k^{\rm HC}) - \mH(\eta) G_k(\eta,\eta_k^{\rm HC}) \right] \nn
\end{eqnarray}
Our next step is to change the integration variable in Eq. \eqref{main5}. In particular, we use that $k(\eta) = m_p a(\eta)$, so $k^2 dk = m_p^3 a(\eta)^2 a'(\eta) d\eta = m_p^3 a(\eta)^3 \mH(\eta) d\eta$. Therefore, after that change of variable, Eq. \eqref{main5} is
\begin{eqnarray}\label{main6}
	\avg{a^{-1} \xi \dvarphi '(x)} &=& \frac{2}{\pi } \int d\eta \: \delta(\eta-\eta_k^{\rm HC}) m_p^3 \mH(\eta)a(\eta_k^{\rm HC})^2  P_{\Delta_k} \nn
	&\times&   \left[ G_k'(\eta,\eta_k^{\rm HC}) - \mH(\eta) G_k(\eta,\eta_k^{\rm HC}) \right] \nn
	&=& \frac{2}{\pi } m_p^3   \mH(\eta_k^{\rm HC})a(\eta_k^{\rm HC})^2  P_{\Delta_k},
\end{eqnarray}
where in the last line we have used that $G(\eta,\eta) = 0$ and $G'(\eta,\eta) =1$; these values are exact and can be directly obtain from Green's function \eqref{greenvk}. Furthermore, given that  $ k = a(\eta_k^{\rm HC})m_p$ and thus $\mH(\eta_k^{\rm HC})=k H(\eta_k^{\rm HC})/m_p$, Eq. \eqref{main6} results in 
\begin{equation}
		\avg{a^{-1} \xi \dvarphi '(x)} = \frac{2}{\pi }  k^3 H(\eta_k^{\rm HC}) P_{\Delta_k}.
\end{equation}

Finally, employing the right-hand-side of Eq. \eqref{main1}, we obtain the power spectrum of $\Delta_{\nk}$, i.e. 
\begin{equation}\label{PSDeltak}
	P_{\Delta_k} = \frac{\pi\gamma}{2  } \frac{H^4 (\eta_k^{\rm HC}) }{k^3} 
\end{equation}

This is the main result of the present subsection. Additionally, we note that Eq. \eqref{PSDeltak} is analogous to the important result of Eq. \eqref{desired} (in the example considered in Section \ref{subsecDS}), which was based on an exact de Sitter spacetime.

\subsection{Power Spectrum of the comoving curvature perturbation $\mR$}

We are now finally in position to calculate the primordial spectrum. The key aspect will be to relate the power spectrum of $\Delta_{\vec k}$, Eq. \eqref{PSDeltak}, with the primordial spectrum. 

The \textit{dimensionless} power spectrum of $\mR$ is defined as
\begin{equation}
\mP_\mR (k) \delta^3 ({\nk}+{\nq}) \equiv \frac{k^3}{2 \pi^2} \avg{\mR_{\nk} \mR_{\nq} }.
\end{equation}
Also, the comoving curvature perturbation $\mR$ is related to the  MS variable as $\mR = v/z$.  Consequently, the primordial power spectrum of $\mR_k$ is given as 
\begin{equation}\label{PSRk0}
	\mP_\mR (k) \delta^3 ({\nk}+{\nq}) = \frac{k^3}{2 \pi^2 z^2} \avg{v_{\nk} v_{\nq} }.
\end{equation}
Here, we note that the primordial power spectrum is traditionally expressed in the super-Hubble regime characterized by $k/aH \to 0$, i.e., when the proper wavelength of the modes exceeds the Hubble radius. In this regime, the amplitude of the primordial curvature perturbation $\mR_k$ remains approximately constant. Hence, our next task is to compute $\avg{v_{\nk}(\eta) v_{\nq}^(\eta)}/z^2$, in the limit $k/aH\to 0$.

Employing the solution shown in Eq. \eqref{solvk} and the definition $z \equiv \sqrt{2 \epsilon_1} M_P a$, we have
\begin{eqnarray}\label{PSvk}
\frac{1}{z^2}	\avg{v_{\nk} (\eta) v_{\nq}(\eta)} &=& \frac{1}{2 \epsilon_1 M_P^2 a(\eta)^2} \int_{-T}^ \eta d\tau_1 \int_{-T}^ \eta d\tau_2 \nn
&\times& G_k (\eta,\tau_1) G_q (\eta,\tau_2) a^2 (\tau_1) a^2 (\tau_2) \nn
&\times& \avg{\Delta_{\nk} \Delta_{\nq}} \delta(\tau_1-\eta_k^{\rm HC}) \delta(\tau_2-\eta_q^{\rm HC})  \nn
&=&  \frac{1}{2 \epsilon_1 M_P^2 a(\eta)^2} \int_{-T}^ \eta d\tau_1 \int_{-T}^ \eta d\tau_2 \nn
&\times& G_k (\eta,\tau_1) G_q (\eta,\tau_2) a^2 (\tau_1) a^2 (\tau_2) \nn
&\times& P_{\Delta_k} \delta^3 ({\nk}+{\nq})\delta(\tau_1-\eta_k^{\rm HC}) \delta(\tau_2-\eta_q^{\rm HC}) \nn
&=& \frac{1}{2 \epsilon_1 M_P^2} P_{\Delta_k} I_G (k,\eta)^2 \delta^3 ({\nk}+{\nq}) 
\end{eqnarray}
where in the second line we have used the definition of $P_{\Delta_k} $, and in the third line we have defined
\begin{eqnarray}\label{IG}
	I_G (k,\eta) &\equiv& \frac{1}{a(\eta)} \int_{-T}^ \eta d\tau_1 \:  G_k (\eta,\tau_1)  a^2 (\tau_1) \delta(\tau_1-\eta_k^{\rm HC}) \nn
	&=& \frac{G_k (\eta,\eta_k^{\rm HC})}{a(\eta)} a(\eta_k^{\rm HC})^2. 
\end{eqnarray}

Consequently, comparing Eqs. \eqref{PSvk} and \eqref{PSRk0}, and using Eq. \eqref{IG}, we obtain the dimensionless spectrum
\begin{equation}\label{PSRk1}
	\mP_\mR(k,\eta) = \frac{k^3}{2 \pi^2}\frac{P_{\Delta_k}}{2 \epsilon_1 M_P^2} \frac{G_k (\eta,\eta_k^{\rm HC})^2}{a(\eta)^2} a(\eta_k^{\rm HC})^4.
\end{equation}
Thus, we note that the dependence on the spectrum $P_{\Delta_k}$ has become manifest in the expression for the primordial power spectrum, Eq. \eqref{PSRk1}. 

We now employ the Green's function \eqref{greenvk} to approximate $G_k (\eta,\eta_k^{\rm HC})^2/a(\eta)^2$ [in Eq. \eqref{PSRk1}] by taking the late time limit $-k \eta \to 0$, i.e.
 \begin{eqnarray} \label{PSRk2}
 	\mP_\mR(k,\eta)\bigg|_{-k \eta \to 0} &\equiv& \mP_\mR(k) \nn 
 	&\simeq& \frac{k^3}{2 \pi^2} \frac{P_{\Delta_k}}{2 \epsilon_1 M_P^2} a(\eta_k^{\rm HC})^4  \bigg( \frac{H^2 \pi^2}{4 k^4} \bigg) \nn 
 	&\times& 2^{2\nu} J_\nu (1)^2 \Gamma(\nu)^2 (-k \eta)^{3-2\nu},
 \end{eqnarray}
where $\Gamma(\nu)$ is the Gamma function, and recall that $\nu = 3/2 + \epsilon_1 + \epsilon_2/2$.\footnote{The peculiar function $J_\nu (1)$ appears because one of the arguments of the Green's functions is evaluated at $\eta_k^{\rm HC}$, namely at the time of horizon crossing. In particular, when $k = a(\eta_k^{\rm HC}) m_p= - m_p / H( \eta_k^{\rm HC} )  \eta_k^{\rm HC} $, this implies $ -k \eta_k^{\rm HC} = 1 $ because $H( \eta_k^{\rm HC} ) \simeq m_p$.  So the function $J_\nu (-k \eta_k^{\rm HC}) \simeq J_\nu (1).$} Moreover, using the expression found for $P_{\Delta_{\nk}}$ \eqref{PSDeltak} and that $a(\eta_k^{\rm HC}) = k/H( \eta_k^{\rm HC} ) $, equation \eqref{PSRk2} becomes
\begin{equation} \label{PSRk3}
	\mP_\mR(k) \simeq \frac{\pi \gamma J_\nu(1)^2 \Gamma(\nu)^2}{4  M_P^2 } \frac{H^2}{\epsilon_1} (-k \eta)^{3-2\nu}.
\end{equation} 
The spectrum \eqref{PSRk3} is approximately constant, that is, it can be shown that 
\begin{equation}
	\frac{d}{d \eta} \bigg[ \frac{H(\eta)^2}{\epsilon_1(\eta)} (-k \eta)^{3-2\nu} \bigg] = \mathcal{O}(\epsilon_i^2).
\end{equation}

Given the previous result, it is customary to evaluate the right hand side of Eq. \eqref{PSRk3} at a time $-\eta_p =1/k_p$ this is, at the time of horizon crossing for a pivot mode $k_p$. Therefore, Eq. \eqref{PSRk3} yields,
\begin{equation} \label{PSRk4}
		\mP_\mR(k) \simeq \frac{\pi J_{\nu}(1)^2 \Gamma(\nu)^2}{4 } \frac{\gamma H_\diamond^2}{M_P^2 \epsilon_{1 \diamond}} \bigg( \frac{k}{k_p} \bigg)^{3-2\nu},
\end{equation}
where the notation $\diamond$ indicates that the variables are evaluated at the time $-\eta_p =1/k_p$. 
From expression  \eqref{PSRk4}, we can obtain the prediction for the inflationary parameters, namely the spectrum's amplitude $A_s$ and the scalar spectral index $n_s$. Specifically, the amplitude is given by
\begin{equation}\label{num1}
	A_s = \mP_\mR(k_p) \simeq \frac{\pi^2 J_{3/2}(1)^2}{16} \frac{\gamma H_\diamond^2}{M_P^2 \epsilon_{1\diamond}},
\end{equation}
where  we have evaluated the Gamma and Bessel functions at $\nu \simeq 3/2$ (in fact $\Gamma(3/2)^2 = \pi/4$)

It is instructive to compare this prediction with the standard result from slow-roll inflation, namely
\begin{equation}\label{Asstd}
	A_s^{\rm std} = \frac{H_\diamond^2}{8\pi^2 M_P^2 \epsilon_{1\diamond}}.
\end{equation}

In each case, a numerical prefactor appears: in our approach it is $\pi^2 J_{3/2}(1)^2 / 16$, whereas in the standard scenario it is $1/(8\pi^2)$. Clearly, they are not equal, but in both instances, this factor is of order $10^{-2}$. The algebraic difference in these prefactors arises from the choice of initial conditions for the field modes. In the standard case, the initial conditions are set by the adiabatic  vacuum, which evaluates the field modes at past infinity ($\eta \to -\infty$), whereas in our approach they are imposed at the time of horizon crossing ($-k \eta_k^{\rm HC} = 1$).

As a consequence, in our model, if $H_\diamond \simeq m_p$, then one requires 
\begin{equation}\label{estgamma}
\gamma \simeq 10^{-9} \epsilon_{1 \diamond}
\end{equation}
for the amplitude to be consistent with the observed CMB temperature anisotropies. This contrasts with the standard prediction, in which $H_\diamond$ must lie several orders of magnitude below the Planck mass to match the CMB amplitude, which then gives rise to the trans-Planckian problem mentioned in the Introduction.

The spectral index, on the other hand, is defined as  
\begin{equation}\label{num2}
	n_s - 1 \equiv 	\frac{d \ln \mP_\mR(k)  }{d \ln k} \bigg|_{k=k_p}.
\end{equation}
Therefore, the spectral index obtained from Eq. \eqref{PSRk4} and \eqref{nuni} is
\begin{equation}\label{nsfinal}
n_s  = 1 -2 \epsilon_{1 \diamond} - \epsilon_{2 \diamond}.
\end{equation}
The latter prediction is the same as the usual one from slow-roll inflation.

It is tempting to confront the model to the present data and obtain an estimate for the numerical values of $\epsilon_{1 \diamond}$ and $\epsilon_{2 \diamond}$. If we associate $\gamma=m_p^{-2}\partial_{\varphi \varphi} V$, we are able to impose two observational constraints: $\gamma =10^{-9}\epsilon_1$, which fixes the amplitude of the power spectrum, and $1-n_s=4\times 10^{-2}$, the scalar spectral index—under the assumption that the $\epsilon_i$ are small so that we may work to first order. Additionally, the derivatives of the potential can be related to the HFF \cite{Leach:2002ar}; in particular $m_p^2\partial_{\varphi \varphi} V/V= 2\epsilon_1-\epsilon_2/2$. As a consequence, we have
	\ba\label{estim1}
	\gamma=m_p^{-2}\partial_{\varphi \varphi} V=2\epsilon_1-\frac{\epsilon_2}{2} =10^{-9} \epsilon_1 \nn
	1-n_s=2\epsilon_1+\epsilon_2=4\times 10^{-2},
	\ea
	where we have used Eqs. \eqref{estgamma} and \eqref{nsfinal}. The solution is
\be\label{estim2}
\epsilon_1\approx 7 \times 10^{-3} \qquad \epsilon_2=3 \times 10^{-3}.
\ee
Being massless, for tensor modes we expect $\gamma_{\rm tensor}=0$.
Note that the mass scale that fits the data is $m/m_p=\sqrt {\gamma}\approx 10^{-5}$ is not far from GUT scale in some models. Thus, even when the scale of inflation is the Planck scale in this scenario, the GUT scale could arise here as well in the mass of the (scale invariance breaking) scalar degree of freedom (see discussion in Section \ref{sec_4})
\footnote{The assumption that the scale of inflation is close to the Planck scale can potentially lead to a reheating phase at an energy scale higher than the GUT scale (see \cite{Bengochea:2024msf}) which might seem to conflict with one of the attractive features of inflation: the dilution of the magnetic monopoles possibly produced during the GUT phase transition. This conclusion is, of course, contingent upon the validity of the GUT framework, which, despite its elegant formulation, depends on multiple assumptions. In particular, existing limits on proton decay tend to support supersymmetric extensions or more sophisticated model arrangements that incorporate further nontrivial assumptions.}.

\section{Tensor modes}\label{sec_4}

In this section we argue that gravitational waves are not produced directly by the fundamental discreteness.
However, tensor modes would still be produced at next-to-leading order due to the previously described scalar modes.

An intriguing property of our universe is its time asymmetry, which is evident in the validity of the second law of thermodynamics. The macroscopic irreversibility of the universe’s trajectory in phase space highlights the extraordinarily special nature of its primordial state as it emerges from what we classically refer to as the big bang singularity.
As Penrose emphasized \cite{penrose1989difficulties}, the extreme fine-tuning required for the initial state to produce the universe we observe is not satisfactorily explained by the inflationary hypothesis. He suggests that a more fundamental explanation must be found within a theory of quantum gravity and formulates this idea through the \textit {Weyl curvature hypothesis}, which posits that the big bang singularity must give rise to a geometry with initially vanishing Weyl curvature.

The fine-tuning issue highlighted by Penrose arises from the enormous increase in the universe’s entropy, $S_{\rm U}$, since the big bang. This growth is primarily driven by gravity during the universe’s dynamical evolution, with black hole entropy---resulting from gravitational collapse---being the dominant contributor.
In some approaches to quantum gravity, such as loop quantum gravity, this vast entropy is explained by the fundamental granularity of spacetime \cite{Perez:2017cmj}. In this framework, black hole entropy corresponds to the counting of all microscopic states of the discrete geometry of the black hole horizon, consistent with the macroscopic coarse-grained description provided by general relativity.
For a universe like ours to allow for the observed entropy increase, its initial conditions must be fine-tuned among $\exp{S_{\rm U}}\gg 10^{10^{123}}$ different states—as Penrose argues for our visible universe.

The perspective that black hole entropy originates from the fundamental granularity of geometry at the Planck scale also provides a natural resolution to another puzzle related to the second law: Hawking's information paradox.
According to this view \cite{Perez:2014xca}, at the end of black hole evaporation, the microscopic granular structure of spacetime geometry is responsible for the purification of Hawking radiation. The final state of black hole evaporation—a pure state, as expected from unitarity—contains microscopic Planckian degrees of freedom entangled with the long-wavelength (IR) radiation degrees of freedom in the Hawking quanta.
This entanglement is generated in the strong quantum gravity region classically identified as the singularity, which, unlike the initial big bang singularity, would be extremely disordered. From the perspective of a coarse-grained observer, who is insensitive to Planck-scale granularity, entropy continues to increase due to the growingly chaotic nature of the microscopic structure of spacetime quantum geometry making the fine tuning issue raised by Penrose even stronger.

Thus, Penrose's paradox translates into the question of why the universe was born in such an ordered, low-entropy microscopic state. Here, we do not attempt to answer this question but rather take it as a necessary condition—an implicit assumption in the usual approach, even when not always explicitly emphasized. We assume that the background geometry and matter fields are in an extremely special state. The state is consistent with the FLRW ansatz at long distances and 
is as symmetric as discreteness allows down to the Planck scale: a sort of minimal entropy ground state. As a result of the intrinsic granularity,  modes of the scalar field are excited randomly away from the symmetric configuration as they emerge at super-horizon (super-Planckian) scales in semiclassical states with negligible quantum fluctuations from the mean
(which is allowed by the extreme damping of these modes as soon as they are born). Such a state is highly peculiar from the usual quantum field theory perspective, in that it is excited away from the symmetric vacuum as well as strongly squeezed (peaked in geometric discreteness and spread in the conjugate extrinsic curvature). However, as emphasized in the introduction, given the underlying Planckian discreteness, it is actually the adiabatic vacuum that loses its naturalness. From this perspective, the generation of scalar inhomogeneities corresponds to the first dissipative effect where entropy starts growing away from the primordial low entropy big bang.

However, the discreteness hypothesis is a subtle one, and a naïve interpretation can lead to inconsistencies. One common misconception is the idea that quantum gravity could induce violations of Lorentz invariance in the low-curvature regime \cite{Mattingly:2005re}. However, it has been shown that such effects cannot remain small---in the sense of being suppressed by some power of the energy of the process in Planck units, $E/m_p$---as required by observations, unless extreme fine-tuning at the fundamental scale is involved \cite{Collins:2004bp}. This is because loop corrections generally introduce terms in the effective action that include unsuppressed Lorentz-violating operators of dimension lower than four. This theoretical result is strongly supported by a wide range of observational data, which has placed stringent constraints on the parameters governing phenomenological corrections \cite{Kostelecky:2008ts}.

How should one then understand the discreteness involved in the present discussion? Fundamental approaches predicting discreteness face the challenge of explaining how the underlying discreteness of geometric operators is compatible with the absence of preferred frames \cite{Rovelli:2002vp}. In this regard, a consistent perspective emerging from background-independent approaches to quantum gravity is that observables in quantum gravity correspond to diffeomorphism-invariant quantities (Dirac observables), defined through relational descriptions that resolve the constraints in the Hamiltonian formulation \cite{Rovelli:2001bz}.

At the physical level, this idea closely relates to Einstein's view of spacetime geometry as determined by the synchronization of idealized clocks within a properly arranged system of measuring rods. A form of granularity compatible with the absence of background structures would therefore be one defined relationally through the interaction between matter and geometry (the one that physical clocks and rods could probe). Whatever the fundamental description of such degrees of freedom may be, one expects that massless degrees of freedom, in the emergent effective description, cannot interact directly with Planckian discreteness, as their intrinsic scale invariance prevents them from functioning as clocks or rods. In other words a massless excitation cannot define by itself a preferred with respect to which one could give an operational meaning of the Planck scale.

This is in contrast with the scalar field of the previous discussion which breaks scale invariance.  Just as the mass of a field—an energy scale that might initially seem in tension with Lorentz invariance in the absence of gravity—must be understood as the energy of a field excitation in the rest frame defined by its excitations, the formalism of canonical quantum gravity strongly suggests that Planckian discreteness must also be understood in similar terms. That is, it should be apparent only for those degrees of freedom capable of selecting their own dynamical rest frame, where the Planck scale would acquire a meaning. This reasoning singles out the (non-conformally coupled) scalar field, in effective descriptions of inflation, as the primary mediator of this interaction, thereby inheriting the inhomogeneities present at the Planck scale through the mechanism described in this paper.\footnote{Photons and other Standard Model degrees of freedom are scale-invariant. We extend this assumption to the possibly yet unknown degrees of freedom in the matter sector. In this context, it is worth noting that while conformal invariance may hold for these degrees of freedom at tree level, it is likely to be broken by quantum fluctuations, leading to the well-known trace anomaly. However, we emphasize that the trace anomaly is state-independent and thus cannot serve as an indicator of the sensitivity of excitations (away from the vacuum) to the fundamental scale.} At the end of the previous section we have seen how natural values for the slow roll parameters explaining observations follow from the identification of the `diffusion' coefficient $\gamma$ with the effective mass of the scalar field.

However, tensor modes are massless and correspond in addition to excitations of the Weyl curvature, which—according to the Weyl curvature hypothesis—is assumed to vanish asymptotically back toward the big bang singularity \footnote{The vanishing of tensor modes is equivalent to the vanishing of the Weyl tensor only asymptotically to the past if we assume that the initial phase is exactly de Sitter-like. Contributions from matter (in particular, from scalar inhomogeneities) are non-vanishing when the Hubble rate begins to change \cite{Frob:2014fqa}.}. The relational perspective on the nature of Planckian granularity, combined with the fact that Weyl curvature is manifestly scale-invariant, justifies a strong form of the Weyl curvature hypothesis where tensor modes are born at super-horizon scales in a semiclassical state (similar to the scalar field modes in their semiclassical features) but with vanishing expectation values for the tensor metric perturbations \footnote{As in our estimates of \eqref{estim1} and \eqref{estim2}, we could expect $\gamma_{\rm species}$ in \eqref{ghhhhh} to be proportional to $m^2_{\rm species}/m_p^2$. Such is the result in certain proposals based on non-standard assumptions in QFT \cite{Agullo:2008ka}.}. Tensor modes will certainly be produced at next-to-leading order as a consequence of the appearance of scalar modes. Their suppressed power spectrum will be computed in a separate work \cite{gabrieles}.

\section{Discussion and Conclusions}\label{discu}

The standard inflationary paradigm makes explicit assumptions about the initial state of the universe when choosing the FLRW description of the background geometry, assuming an initial de Sitter configuration, and selecting the quantum state of perturbations to be the Bunch-Davies vacuum (or an equivalent asymptotic Hadamard state, which is a trans-Planckian hypothesis). These choices are not explained by the formalism, nor are they implied by the inflationary dynamics, as is sometimes misunderstood.

In quantum field theory on curved spacetimes and the semiclassical approach to quantum gravity, one demands that acceptable quantum states resemble the vacuum in the short-wavelength regime (the Hadamard condition).  This requirement is necessary for the renormalization of the expectation value of the energy-momentum tensor. While this selection criterion is reasonable away from strong-curvature regimes, it generally fails when approaching singularities (for instance, in all reasonable quantum states considered in the description of Hawking radiation by black holes).

So, why would this assumption be natural as one approaches the initial Big Bang singularity? Why should the preferred quantum state (as postulated in standard inflationary cosmology) be a vacuum (FLRW symmetric)  state? If this choice is correct, it clearly requires justification from a more fundamental description beyond the standard inflationary framework. We emphasize that the classical mechanisms by which inflation is claimed to resolve the horizon and flatness problems do not explain the standard selection of the inflaton quantum state. Why should the universe emerge from a singularity in such a specific quantum state? This question arguably represents the most striking manifestation of the trans-Planckian problem. 

In this paper, we propose an alternative primordial quantum state, which seems more natural from the perspective of some approaches to quantum gravity, but would also require proper justification  within a suitable framework of quantum gravity. We assume that the very special nature of the initial state resides in its extremely low entropy and postulate that it deviates strongly from the vacuum state by breaking the FLRW symmetries for scalar modes as soon as these modes emerge from the Planck scale and become super-Hubble (in a way, as soon as we are able to describe them as quantum field theory excitations on a background geometry). A consistency requirement is that the Hubble rate during the de Sitter phase is Planckian. We have shown that this leads to a power spectrum of curvature perturbations that relates to the observations in the CMB in a way that is similar but not equal to the standard formulation when parametrized by slow-roll parameters. The model predicts a primordial spectrum with the correct shape and amplitude. However, unlike the standard approach, our formulation does not involve the symmetry-breaking issues related to the quantum-to-classical transition: symmetry is broken from the onset in our formulation.

The gravitational degrees of freedom can be separated into Ricci-type components (which violate scale invariance and are fully determined via Einstein's equation by the matter content) and those encoded in the Weyl tensor (the purely gravitational degrees of freedom), which are scale-invariant. The Weyl curvature hypothesis sets these to zero in the background geometry, while the relational hypothesis about the nature of discreteness demands that no tensor mode would be excited by the granularity. The stochastic mechanism generating the scalar modes does not operate on such scale-invariant degrees of freedom; gravitational waves are only produced at next-to-leading order by the scalar inhomogeneities acting as sources. This framework thus permits a Planckian Hubble rate without running into any obvious conflict with the observed low value of the tensor-to-scalar ratio in the CMB.

An initially large-Hubble-rate inflationary phase (with $H \approx m_p$) eventually slows down during the reheating epoch, where the universe would reach temperatures of order $m_p$ if the process is sufficiently rapid; this leads to phenomenology---including a possible mechanism for dark matter genesis---that remains robust despite uncertainties in the detailed matter content at such scales \cite{Bengochea:2024msf}. As the Hubble rate decreases away from the Planck scale (entering a radiation-dominated regime), the interaction mechanism between the microscopic granularity and the relevant field modes ceases and the quantum states of all degrees of freedom become Hadamard (look like the Minkowski vacuum in the UV). Understanding how such regulated behavior of quantum field modes emerges from the microscopic quantum‐gravity realm is a central question for approaches to quantum gravity that predict discreteness at the Planck scale.

This question is also intimately connected with the emergence of Lorentz invariance from such quantum gravity formulations. From the viewpoint offered by this work, the issue of Lorentz invariance appears closely linked to entropy and the fine-tuned nature of the universe's initial conditions. As discussed, the second law of thermodynamics implies that the microscopic architecture of the primordial state must be extremely special. Perhaps among these special features are those that enable the emergence of Lorentz-invariant physics in quantum field theory, particularly in low-curvature (and, most importantly, \textit{sufficiently early}) regimes\footnote{By \textit{early} we mean before a substantial part of the matter content of the universe undergoes gravitational collapse to form black holes that completely evaporate away via Hawking radiation. We are in this early phase in the present state of the universe.}. Indeed, if the scenario put forward in \cite{Perez:2014xca} is correct, spacetime ``grows old'' as black holes form and subsequently evaporate. Once their Hawking radiation has radiated out to infinity, the microscopic degrees of freedom that purify the radiation (retaining the memory of the previously existing black hole) break Lorentz symmetry in a microscopic sense and may become observable  despite the large-scale flatness of spacetime as probed by coarse observers. As suggested by \cite{Collins:2004bp}, this could occur via the emergence of Lorentz-violating dimension-four or lower operators in quantum field theory processes. Consequently, Lorentz invariance in special relativity and standard quantum field theory may be characteristic only of a sufficiently young universe, such as the one we inhabit.

\begin{acknowledgments}

We would like to thank Federico Piazza and Daniel Sudarsky for instructive interactions. G.R.B. is supported by CONICET (Argentina) and he acknowledges support from grant PIP 112-2021-0100225-CO of CONICET (Argentina). G.L. is supported by grants Universidad Nacional de La Plata I+D G175, and PIP 112-2020-0100729-CO of CONICET, and CONICET (Argentina). A.P. aknowledges support of the ID\#62312 grant from the John Templeton Foundation via the QISS Project (\href{qiss.fr}{qiss.fr}).

\end{acknowledgments}

\bibliography{bibliografia}

\end{document}